\documentclass[aps,twocolumn,showpacs,amsmath,amssymb,pra,superscriptaddress,floatfix,longbibliography]{revtex4-1}
\usepackage{braket}
\usepackage{amsmath}
\usepackage{amssymb}
\usepackage{mathtools}
\usepackage{graphicx}
\usepackage{dcolumn}
\usepackage{bm}
\usepackage{multirow}
\usepackage{leftidx}
\usepackage{color}
\usepackage{float}

\usepackage{amsfonts}
\usepackage{eufrak}
\usepackage[german,english]{babel}

\begin{document}
\title{Photon scattering from a cold, Gaussian atom cloud}
\author{F.~Robicheaux}
\email{robichf@purdue.edu}
\affiliation{Department of Physics and Astronomy, Purdue University, West Lafayette,
Indiana 47907, USA}
\affiliation{Purdue Quantum Science and Engineering Institute, Purdue
University, West Lafayette, Indiana 47907, USA}
\author{R.~T.~Sutherland}
\email{sutherland11@llnl.gov}
\affiliation{Physics Division, Physical and Life Sciences, Lawrence Livermore
National Laboratory, Livermore, California 94550, USA}

\date{\today}

\begin{abstract}
We study the effect of a weakly driven atomic cloud's polarization 
distribution on its photon 
scattering lineshape. 
In doing this, we find three distinct polarization regimes. First, for
dilute clouds, the 
polarization magnitude is relatively constant. Second, for denser 
clouds, polarization builds at the front of the cloud for near-resonant 
light. Third, when the cloud condenses to the point where its 
dimensions become comparable to the wavelength, light 
refocuses towards the back of the cloud for red
detuning. For these regimes, we show 
which `mean-field' frameworks accurately describe the differing photon
scattering
lineshapes. Finally, for even denser clouds, mean field models become 
inaccurate and necessitate the full point dipole model that includes 
atom-atom correlations.
\end{abstract}

\maketitle
\section{Introduction}

Scientists study light-matter interactions with an
ostensibly diverse set of physical models. ``Microscopic" models\textemdash 
that treat atoms as point dipoles\textemdash have 
been integral to understanding effects such as superradiant
spontaneous emission \cite{dicke1954,eberly_1971,gross_1982,sutherland_2017_3}, coherent
scattered radiation 
\cite{eberly_1971,ruostekoski1997,scully_2006,scully2007,sutherland_2016_1,sutherland_2016_2,bromley_2016,bettles_2016,svidzinsky_chang2008,ruostekoski_2012,sutherland_2017_4,zhu_2016}, 
collective Lamb-shifts \cite{meir2014}, and Anderson localization
\cite{anderson_1958,storzer_2006,skipetrov_2015}. Within the microscopic
model, some analyses leverage other mean-field approximations such
as assuming an evenly excited phase distribution (timed-Dicke state)
\cite{eberly_1971,scully_2006}, or using clouds that are denser but 
have less atoms than the represented cloud \cite{bromley_2016}. ``Mean-field" 
approaches\textemdash that treat illuminated matter as continuous
dielectrics \cite{JDJ}\textemdash are also an intuitive tool 
for understanding many of these same effects
\cite{mie_1908,friedberg1973,kaiser_2018}.
While there has been recent work towards understanding the relationships
between these treatments \cite{javanainen2014,javanainen_2016,kaiser_2018}, where and why
models fail to reproduce the results of the full point dipole
system remains largely unexplored.

\begin{figure}
\resizebox{80mm}{!}{\includegraphics{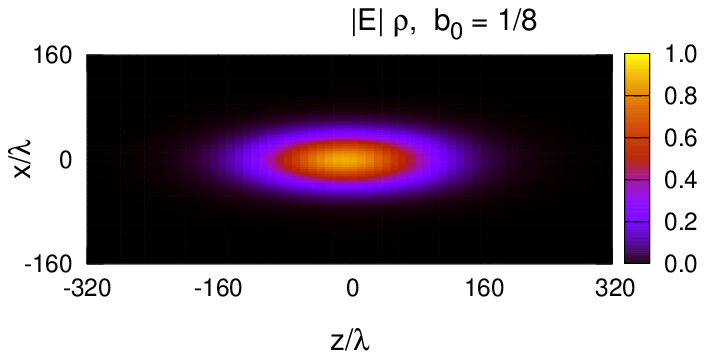}\includegraphics{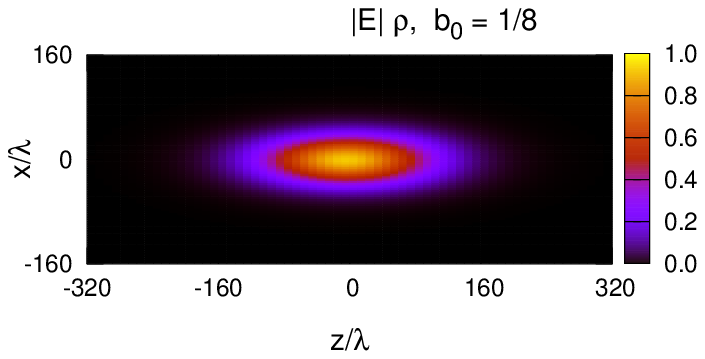}}
\resizebox{80mm}{!}{\includegraphics{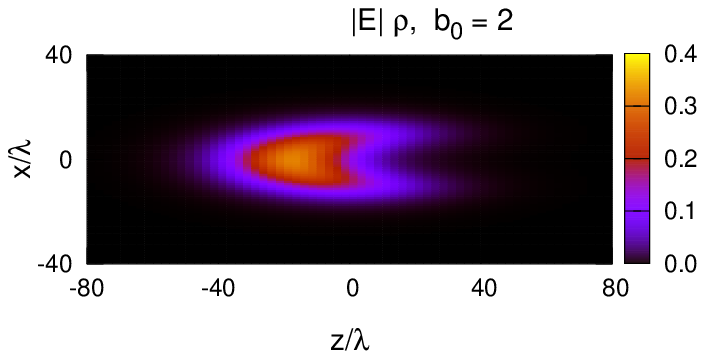}\includegraphics{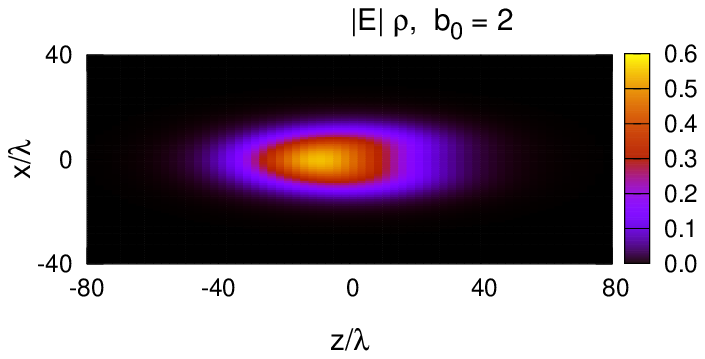}}
\resizebox{80mm}{!}{\includegraphics{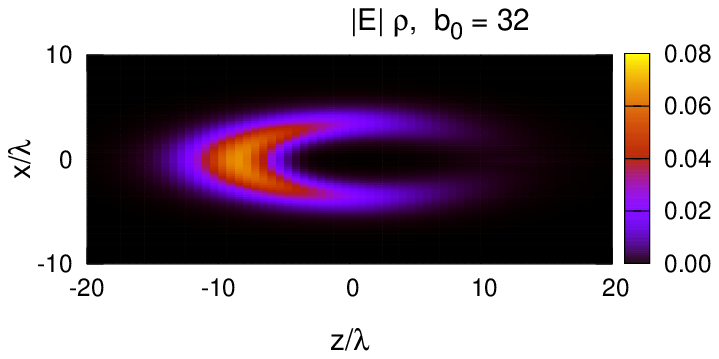}\includegraphics{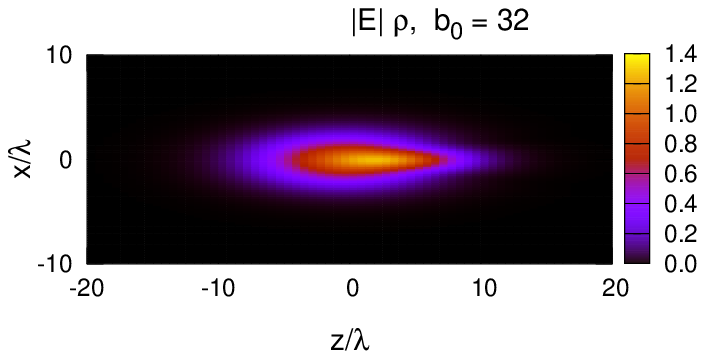}}
\caption{\label{FigSchem}
The continuum model calculation for the $y=0$ cross section of the
spatial dependence of $|E|\rho$ which is proportional to the polarizability.
In all plots, $|E|$ has been
divided by $|E|$ at $z\to -\infty$ and the density is divided by
the peak density; if there were no attenuation or focusing, the $|E|\rho$
would have a maximum value of 1 at $x=z=0$.
In all calculations, the on resonance and on axis optical depths $OD=2b_0$. The
spatial density is characterized by $\xi = 2$, Eq.~(\ref{EqDens})
and $N=2^{11}$. The detuning is set at $\Delta =0$ for the left column
and $\Delta = -0.4(1+b_0/4)\Gamma$ for the right column
to accentuate the focusing effect.
The electric field is scaled by the incident
field and the density is scaled by the maximum density.
}
\end{figure}

In this work, we compare the continuous dielectric model with the full 
point dipole
calculations of a frozen atomic cloud of two-level atoms driven by
a weak laser. In order to do this, we implement a unique iterative
numerical technique capable of simulating over $10^{5}$ atoms without approximation.
We find three distinctive regimes of the
cloud (see Fig.~\ref{FigSchem}), each characterized by its
polarization distribution. The figure shows a continuum dielectric
calculation of the scaled $|E|\rho$ where $|E|$ is
the local magnitude of the electric field and $\rho$ is the average
atom density; in the plots, $y=0$ and the light is polarized in the
$x$-direction and propagates in the $z$-direction.
The left column is for detuning $\Delta =0$
while the right column is for red detuning, $\Delta =-0.4(1+b_0/4)\Gamma$.
First, for clouds with small optical depths, OD, the atom
excitations are nearly evenly distributed throughout the cloud for 
all laser detunings (see 
the $b_0=1/8$ case);
thus, models that assume an even distribution \cite{scully_2006} give
good agreement with the full
point dipole model. When the cloud becomes more dense, the laser
intensity is substantially reduced as it traverses the cloud, causing excitation
to be much more likely in the front of the cloud for near-resonant
light, causing non-Lorentzian lineshapes; this is exemplified by the
$b_0=2$ case where $\Delta =0$, left column, has considerable attenuation
while the detuned case, right column, has more uniform excitation. We find
that, in this regime, our cloud is well described by a continuous
dielectric solved using the eikonal
approximation in Maxwell's equations. When the eikonal approximation 
is valid, the photon
scattering is well described by a single parameter, OD. Finally, for
clouds with dimensions comparable to a wavelength, the light
refocuses towards the back of the cloud for red detuning
\cite{sutherland_2016_1} (see the $b_0=32$ case); the red detuned light
causes larger polarization for $z>0$ through the center of the cloud whereas
there is almost no polarization near the center of the cloud for zero
detuning, causing the
eikonal approximation \textemdash and the cloud's dependence on
OD \textemdash to break down. For all of the cases in Fig.~\ref{FigSchem},
the continuum dielectric model
reproduces the total photon scattering and forward photon scattering
rate from point dipole calculations averaged over many spatial configurations,
even the $b_0=32$ cases although accuracy
requires the higher-order paraxial
approximation, which includes focusing. At even higher densities,
we find differences between the continuum dielectric and point dipole
calculatations; 
the Clausius-Mossotti equations do not improve the agreement between
the point dipole and the continuum model and, in fact, tend to give
worse agreement at higher densities, as was
found in Refs.~\cite{javanainen2014,javanainen_2016}. This is
due to the 
emergence of dipole-dipole correlations between atoms, and
diffraction perpendicular to the laser. 

The paper is organized as follows: Sec.~\ref{SecMeth} contains the
methods used in the calculations, Sec.~\ref{SecRes} contains the
results of the calculations, Sec.~\ref{SecCon} contains a short discussion
of conclusions, and the appendix contains information about the
numerical method used to solve for the coupled dipoles,
Sec.~\ref{SecApp}, and a discussion of the paraxial approximation,
Sec.~\ref{SecPar}. 

\section{Methods}\label{SecMeth}

This section describes the calculation of a plane wave of low intensity 
light interacting
with a Gaussian cloud of atoms. This is done using two separate 
formalisms. In Sec.~\ref{SecAtom},
we describe treating the atoms as stationary and interacting through the point
dipole Green's function \cite{javanainen2014,svidzinsky2010,jenkins2012,rouabah2014,sutherland_2016_1,sutherland_2016_2,bromley_2016,bettles_2016}. 
In Sec.~\ref{SecCont}, we treat the cloud as a continuum dielectric, $\chi_e$, with a Gaussian
spatial dependence. For all calculations we assume that the direction 
of propagation is $z$ and
the direction of polarization is $x$: $\vec{k}=k\hat{e}_z$ and
$\hat{e}_{\rm las}=\hat{e}_x$. All of the calculations assume 
$J=0$ to $J=1$ transitions. Finally,
this section assumes that the change in $k$ over a
resonance line width is much smaller than $k$,
which is an accurate approximation for optical transitions. 

In our calculations, the atoms were given random positions following 
a Gaussian density distribution:
\begin{equation}\label{EqDens}
\rho (x,y,z)=\frac{N}{(2\pi )^{3/2}r_f^3}e^{-[(x^2+y^2)\xi + z^2/\xi^2]/(2r_f^2)}
\end{equation}
where $N$ is the number of atoms, $r_f$ is the geometric mean of the
$x,y,z$ standard deviations of the
atom cloud, $\xi$ is a shape parameter,
and the average density is $\bar{\rho}=N/([4\pi ]^{3/2} r_f^3)$.
The $\xi$ equals 1 for a spherical distribution and
is greater than 1 for a cloud elongated in the direction of propagation.
The cooperativity parameter $b_0 = 3 N/(r_f^2k^2)$ is related to
the on-resonance optical depth, $OD$, through the center of the cloud:
$OD=\xi b_0$.

\subsection{Light scattering from atoms}\label{SecAtom}

In the weak field limit, the effect of a monochromatic beam of wave number,
$\vec{k}$, and polarization, $\hat{e}_{las}$, on a cloud of
atoms can be determined by the $a^{(\alpha )}_j
\equiv\langle \sigma_j^{(\alpha )}\rangle$, i.e. the expectation value of
the lowering
operator for component $j$ of the $\alpha$-th atom \cite{ruostekoski1997,lee_2016,sutherland_2017_4}. 
In this limit,
the equations of motion for the amplitude of oscillation are
\begin{eqnarray}\label{EqPart}
\frac{d a^{(\alpha )}_j}{dt}&=&\left( i\Delta -\frac{\Gamma}{2}\right)a^{(\alpha )}_j - i
\frac{\Omega_R}{2}(\hat{e}_{las}\cdot\hat{e}_j)e^{i\vec{k}\cdot\vec{R}^{(\alpha)}}
\nonumber \\
&\null &-\sum_{\alpha ' \neq \alpha}\sum_{j'}g_{j,j'}(\vec{R}^{(\alpha ,\alpha ')})a_{j'}^{(\alpha ')}
\end{eqnarray}
where $\alpha$ represents an atom index,
the position of atom $\alpha$ is $\vec{R}^{(\alpha )}$, $\vec{R}^{(\alpha ,\alpha ')}=
\vec{R}^{(\alpha )}-\vec{R}^{(\alpha ')}$, and $j,j'$ is indicating the
component.
Here, $\Omega_R$ is the transition Rabi frequency, $\Delta$ is the detuning of the laser
from the transition, and $\Gamma$ is the decay rate of the excited state.
The  $\hat{e}_j$ is the unit vector in the $j$-direction. The point dipole
Green's function $g$ is given by
\begin{equation}
g_{j,j'}(\vec{R})=\frac{\Gamma}{2}\left[\delta_{j,j'}h_0^{(1)}(s)
+\frac{3\hat{R}_j\hat{R}_{j'}-\delta_{j,j'}}{2}h_2^{(1)}(s)
\right]
\end{equation}
where $h_\ell^{(1)}(s)=j_\ell (s)+in_\ell (s)$ are spherical Hankel functions 
of the first kind and
$s=k |\vec{R}|$ \cite{JDJ}. We solve for the steady state 
$\vec{a}^{(\alpha )}$ by
setting the time dependence of Eq.~(\ref{EqPart}) equal to 0 and 
solving the resulting
matrix equation. When the number of atoms, $N$, was small, we numerically
solved the linear equations using standard Lapack programs 
that temporally scale $\propto N^{3}$. When $N$
was larger than $\sim 10^{3}$, however, we solved for $\vec{a}$ using an
efficient, $\propto N^{2}$, iterative
method that we developed. This numerical technique enabled us to simulate
clouds with more than $2\times 10^5$ atoms; the technique is described in the appendix,
Sec.~\ref{SecApp}. For these calculations we used the two state approximation
where only the $\hat{e}_{las}$
component of $\vec{a}$ is nonzero. We compared this to the case where all three
components of $\vec{a}$ are allowed to be non-zero and found only small
changes.

The angular differential photon scattering rate into
$\hat{k}_f$, normalized by $\Omega_R^2/\Gamma$, is given by
\begin{equation}\label{EqSct}
\frac{d\gamma}{d\Omega}=\frac{\Gamma^2}{2\pi\Omega_R^2 N}\left(|\vec{P}(\vec{k}_f)|^2
-|\hat{k}_f\cdot\vec{P}(\vec{k}_f)|^2\right)
\end{equation}
where the $\Omega$ is the solid angle, and
\begin{equation}\label{EqP1}
\vec{P}(\vec{k}_f) \equiv \sum_\alpha \vec{a}^{(\alpha )}e^{-i\vec{k}_f\cdot\vec{R}^{(\alpha )}}.
\end{equation}
Finally, the total scattering rate per atom,
normalized by $\Omega_R^2/\Gamma$, is equal to
\begin{equation}\label{EqSctTot}
\gamma =\int \frac{d\gamma}{d\Omega}d\Omega = -\frac{2\Gamma}{\Omega_R^2 N}\Re [i\frac{\Omega_R}{2}
\hat{e}_{las}\cdot\vec{P}(\vec{k})].
\end{equation}
where $\Re [...]$ means to take the real component.
The dimensionless form of the photon scattering rate, $\gamma$, above 
is useful since the calculations are far from saturation and, thus,
independent of $\Omega_{R}$ up to a scaling factor.

\subsection{Continuum model of photon scattering}\label{SecCont}

To compare to the light scattering
from stationary atoms, we solved Maxwell's equations with a continuum
electric susceptibility, $\chi_e$, giving the equations
\begin{equation}\label{EqMax}
\nabla^2\vec{E}-\vec{\nabla}(\vec{\nabla}\cdot\vec{E})+k^2\vec{E}
=-k^2\chi_e\hat{x}(\hat{x}\cdot\vec{E})
\end{equation}
where the $\hat{x}(\hat{x}\cdot\vec{E})$ on the right hand side accounts
for using the two state
approximation for the transition instead of all three components of $J=1$.

For a $J=0\rightarrow 1$ transition, the low density form of the
electric susceptibility
is\cite{RL1}
\begin{equation}\label{EqChild}
\chi^{(ld)}_e (\Delta ) =\frac{\chi^{(ld)}_e (0)}{1 - (2 i\Delta /\Gamma )}=
\frac{i\rho\sigma /k}{1 - (2 i\Delta /\Gamma )}
\end{equation}
where $\Delta$ is the laser detuning, $\sigma$ is the cross section for
scattering
photons out of the original direction, and $\rho$ is the density of
atoms in Eq.~(\ref{EqDens}).
For a $J=0\rightarrow 1$ transition, the cross section is $\sigma = 6\pi /k^2$.
The electric susceptibility for a perfect, homogeneous, and
isotropic gas is given by the Clausius-Mossotti
(or Lorentz-Lorenz) form\cite{JDJ}
\begin{equation}\label{EqChi}
\frac{\chi_e}{\chi_e + 3}=\frac{1}{3}\chi_e^{(ld)}\quad\Rightarrow\quad
\chi_e (\Delta )= \frac{i\rho\sigma /k}{1 - (2 i \Delta '/\Gamma)}
\end{equation}
where $\Delta '= \Delta +(\Gamma\rho\sigma /[6 k])$. This gives
a density dependent shift of the resonance  of $-\pi\Gamma\rho /k^3$.

In the paraxial approximation \cite{lax1975maxwell}, the scattered wave has a
slow dependence in the direction transverse to $\vec{k}$. Assuming $\vec{k}=k\hat{e}_z$ and the polarization of the
incoming light to be $\hat{e}_x$, the electric field
is approximated by
\begin{equation}\label{EqEf}
\vec{E}(x,y,z)\simeq\hat{e}_xe^{ikz}E_0\psi_x (x,y,z)
\end{equation}
where, to lowest order,
\begin{equation}\label{EqPar}
i\frac{\partial\psi_x}{\partial z}=-\frac{1}{2k}\nabla_T^2\psi_x -\frac{k}{2}\chi_e\psi_x
\end{equation}
with $\psi_x (x,y,z\to -\infty ) = 1$,
$\nabla_T^2=\partial^2/\partial x^2 + \partial^2/\partial y^2$, and
there is spatial dependence to $\chi_e$ from the density, 
Eq.~(\ref{EqDens}). Higher order terms are discussed in the appendix,
Sec.~\ref{SecPar}.
Except for Fig.~\ref{FigTotHD}, the first and second order
corrections did not significantly change
the results, implying the paraxial approximation is accurate for the case
discussed in Sec.~\ref{SecHD}.

We also considered the eikonal approximation,
which simplifies Eq.~(\ref{EqPar}) as
\begin{equation}\label{EqEik}
i\frac{\partial\psi_x}{\partial z}=-\frac{k}{2}\chi_e\psi_x\quad\Rightarrow\quad \psi_x =e^{i\frac{k}{2}
\int_{-\infty}^z\chi_e(x,y,z')dz'}
\end{equation}
which leads to an analytic $\psi_x$. As a result, the detuning dependence of
the forward and total photon scattering is \textit{fully} described by
OD for systems where the eikonal
approximation is accurate. This means that\textemdash in this 
regime \textemdash clouds with larger $N$ and smaller $\rho$ may
be accurately described by clouds with smaller $N$ and larger $\rho$.
 
Using Eq.~(\ref{EqEf}), the amplitude of oscillation for the $\alpha$-th atom is
approximately
\begin{equation}\label{EqMod}
\vec{a}^{(\alpha )}=\hat{e}_x\frac{\Omega_R}{2\Delta +i\Gamma}e^{ikZ^{(\alpha )}}
\psi_x (\vec{R}^{(\alpha)}).
\end{equation}
The sum in Eq.~(\ref{EqP1}) is approximated as an integral
\begin{eqnarray}\label{EqPmod}
\vec{P}(\vec{k}_f)&\simeq&\int \vec{a}(\vec{r})\rho (\vec{r})d^3r \nonumber \\
&=&\hat{e}_x\frac{\Omega_R}{2\Delta +i\Gamma}\int \rho (\vec{r})e^{i (\vec{k}-\vec{k}_f)\cdot \vec{r}}
\psi_x  (\vec{r})d^3 r.
\end{eqnarray}
This expression can be used in Eq.~(\ref{EqSct})
to obtain the differential scattering rate in the forward direction,
or in Eq.~(\ref{EqSctTot}) to obtain the total scattering rate.
However, it can not be used
for angles substantially different from the forward direction
because of the paraxial approximation and
because it does not account for the
random scattering from individual atoms, which dominates at larger angles.

\begin{figure}
\resizebox{80mm}{!}{\includegraphics{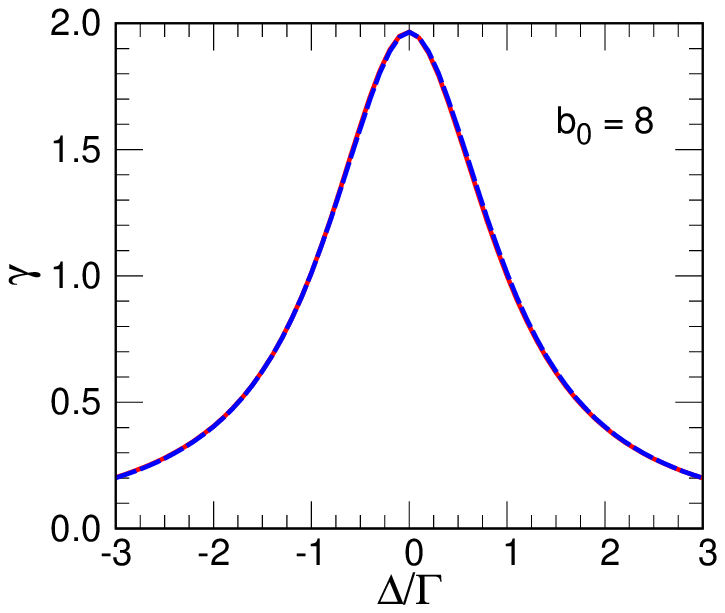}}
\resizebox{80mm}{!}{\includegraphics{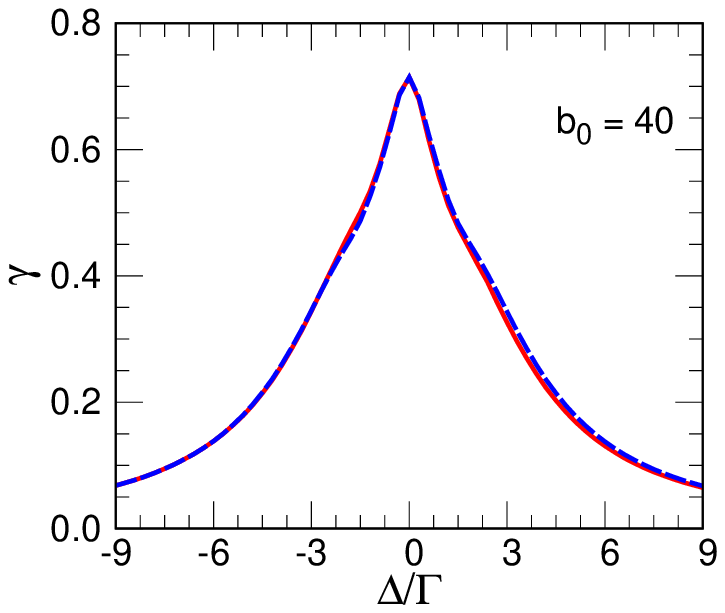}}
\caption{\label{FigTot}
The total scattering rate per atom, Eq.~(\ref{EqSct}), versus
detuning for $\xi =1$ and for different
cooperativity parameters, $b_0$, calculated using
Eq.~(\ref{EqPart}). Each plot shows the results for
two different numbers of atoms: solid (red) line $N=2^{11}$
and long dashed (blue) line $N=2^{17}$. For both plots, the different
calculations are nearly the same, so that the different lines
are nearly indistinguishable. Plots for $b_0=16$, 24 and 32 
and for $N>2^{11}$ showed similar
level of agreement.
}
\end{figure}

The maximum $\chi_e$ is when $\Delta ' =0$ and
$\vec{r}=0$ which leads to $\chi_{e,max}=i b_0^{3/2}/\sqrt{6\pi N}$.
This can be used to
estimate how well the low density limit holds everywhere in
the gas. However, this estimate can be misleading because
the light does not reach the center of the cloud when $b_0$ is
large and diffraction can be
ignored. For light going on axis, the intensity is decreased by the
factor of $\exp (-\xi b_0/2)$ when it reaches the center of the
cloud for the spatially large clouds. However, atom clouds with smaller
$N$ are also spatially small and diffraction becomes increasingly important,
see the $b_0=32$ case of Fig.~\ref{FigSchem}.
We performed calculations for $b_0=8$, 16, 24, 32, and 40 although
only results for 8 and 40 are given below.
The $b_0\geq 24$ cases have negligible intensity at the
center of the cloud if diffraction effects can be ignored. However, for some
of our parameters, diffraction is important and light has
non-negligible intensity at the cloud center for some of the large $b_0$ cases.

\section{Results}\label{SecRes}

We initially present results for the dilute gas limit in order 
to illustrate the effects of absorption and
focusing by the cloud; these calculations require larger $N$
because $\chi_{e,max}$ is inversely proportional to $\sqrt{N}$
for a fixed $b_0$ and $\xi$, which necessitates our iterative numerical
method (see appendix, Sec.~\ref{SecApp}). 
We find that, in this dilute regime, the
continuum model gives excellent agreement with the point dipole
model. We then show that, for dense clouds, calculations based
on classical electrodynamics treatments break down. This indicates
the importance of correlations between neighboring atoms induced
by dipole-dipole interactions.

\subsection{$\xi = 1$, dilute gas limit}

In this and the next
section, all of the calculations of the paraxial approximation of the
continuum use the low density limit of the electric susceptibility,
Eq.~(\ref{EqChild}). This choice is explained in Sec.~\ref{SecHD}.

The first results are for the total scattering per atom for
different number of atoms
for $b_0 = 8$ and  40 and $\xi = 1$. In Ref.~\cite{sutherland_2016_1}, it was shown that the
width of the resonance was $\Gamma '=(1 + b_0\xi /8)\Gamma$ so these
calculations should give resonances from $2\times$ to $6\times$ that of
the single atom resonance when $\xi =1$ and from $3\times$ to $11\times$
larger when $\xi = 2$. Calculations were also done for
$b_0=16$, 24 and 32 but are not reported because their properties
can be inferred
from the calculations described below. Results are reported for $2^{11}=2048$
and $2^{17}=131,072$ atoms.
The calculations were averaged over many runs until a
total of $2^{19}$ atoms were included in the calculation.

To give some rough sizes, the peak density times $\lambda^3$ gives the peak
number inside a cubic wavelength. This is $(b_02\pi /3)^{3/2}/\sqrt{N}$.
For $b_0=40$ and $N=2^{17}$, there are $\sim 2$ atoms per cubic wavelength
at the center of the cloud while $N=2^{11}$ gives 17 atoms per cubic wavelength.
If the more relevant quantity is the density times
$1/k^3$, then the max number is  $(b_0/[6\pi ])^{3/2}/\sqrt{N}$. By this
quantity, all calculations in Fig.~\ref{FigTot}
have much less than 1 atom per $1/k^3$
($\simeq 0.07$ atoms for $b_0=40$ and $N=2^{11}$).

Figure~\ref{FigTot} shows plots of the total scattering rate per atom
versus the detuning
for 2 different cooperativity parameters: $b_0=8$ and 40. In each plot, there
are two calculations for different values of $N$. The $N$ values are chosen
to be
$N=2^{11}$ and $2^{17}$. Despite the drastically different
parameters for each of the clouds, the overall lineshape depends only
on the value of OD. Although $N$ varies by a factor of 64, the total
scattering rate per atom is essentially the same. This is because
for these parameters, the eikonal approximation, Eq.~(\ref{EqEik}),
gives very good agreement
with the calculations from randomly placed atoms. There are two
interesting trends to note. The first is that the resonance line
width is increasing
with $b_0$. This effect was described in
Refs.~\cite{sutherland_2016_1,eberly_1971,scully2007,bromley_2016}. The second is that the line shape
is changing with increasing $b_0$. For $b_0=8$, the line is approximated by
a Lorentzian. However, for $b_0=40$, the
central part of the line is narrower than for a Lorentzian.
Also, the region near $\Delta =0$ appears
to have a dependence like $|\Delta |$, instead of $\Delta^2$, for $b_0=40$.
Calculations were performed for $b_0=16$, 24,
and 32 with similar levels of agreement. 

\begin{figure}
\resizebox{80mm}{!}{\includegraphics{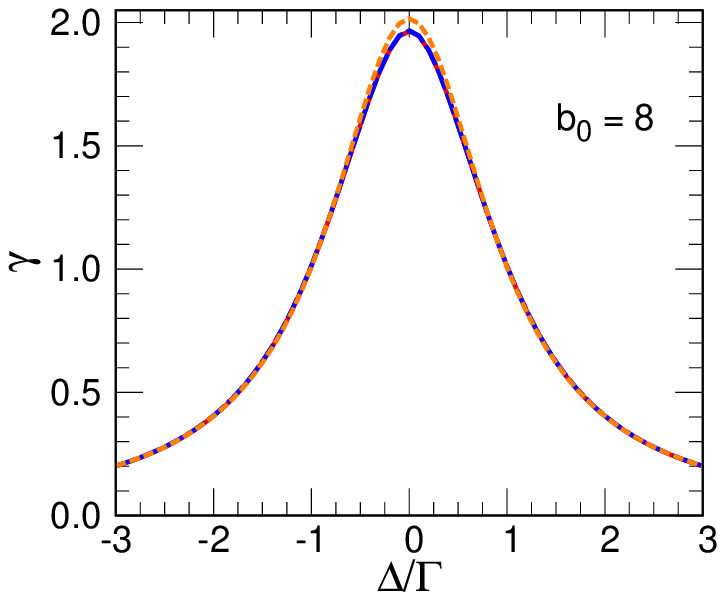}}
\resizebox{80mm}{!}{\includegraphics{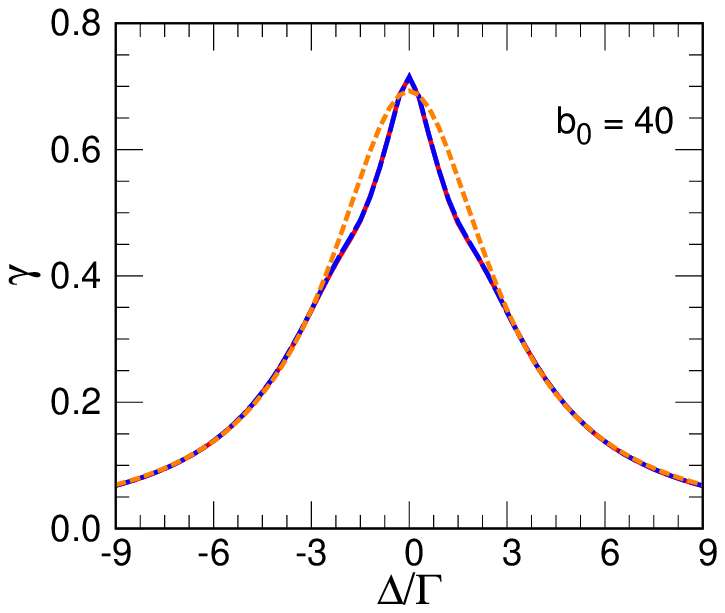}}
\caption{\label{FigTotM}
The total scattering rate per atom, Eq.~(\ref{EqSct}), versus detuning
for $\xi = 1$
and for different
cooperativity parameters, $b_0$. Each plot shows the results for
the full calculation, Eq.~(\ref{EqPart}), with $N=2^{17}$ solid (red) line,
the model calculation, Eq.~(\ref{EqMod}), long dash (blue) line, and
for a Lorentzian proportional to $1/(1+(2\Delta/\Gamma ')^2)$,
dashed (orange) line. For all plots, the model and the full
calculations are in such good agreement that the different lines
are indistinguishable. Plots for $b_0=16$, 24, and 32 showed similar
level of agreement.
}
\end{figure}

Figure~\ref{FigTotM} shows a comparison of 3 different calculations of the
total photon scattering rate, $\gamma$, versus detuning for 2 different coherence
parameters: $b_0=8$ and 40.
The solid (red) line is using the full calculation Eq.~(\ref{EqPart}),
with $N=2^{17}$ atoms in each run and was averaged until $2^{19}$ atoms
were included. The long dash (blue) line is from the paraxial approximation in
Eq.~(\ref{EqMod}) substituted into Eq.~(\ref{EqPmod}). The resulting
$\vec{P}(\vec{k})$ was then used in Eq.~(\ref{EqSctTot}). There were
no adjustable parameters. The model calculation accurately reproduces the full calculation. The dashed (orange) line is a Lorentzian using the
width, $\Gamma ' =(1+b_0\xi /8)\Gamma$, with the height of the Lorentzian
fit to give agreement in the wings. Similar results were found for $b_0=16$, 24, and 32.

\begin{figure}
\resizebox{80mm}{!}{\includegraphics{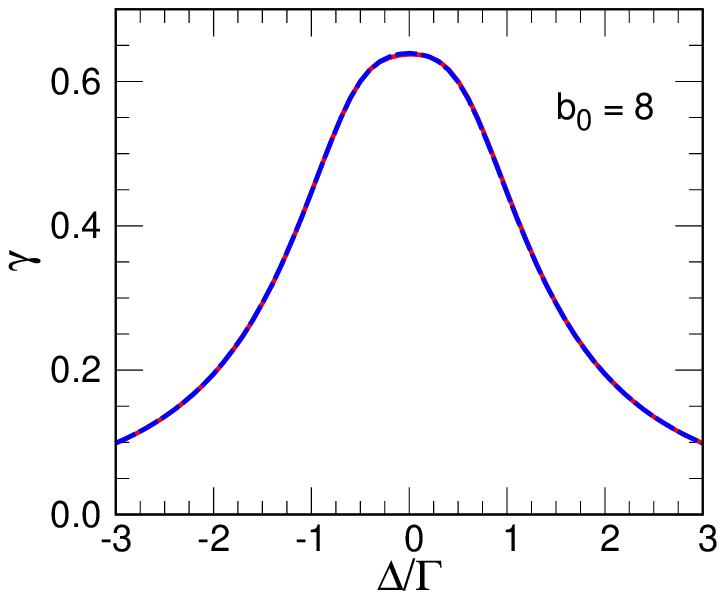}}
\resizebox{80mm}{!}{\includegraphics{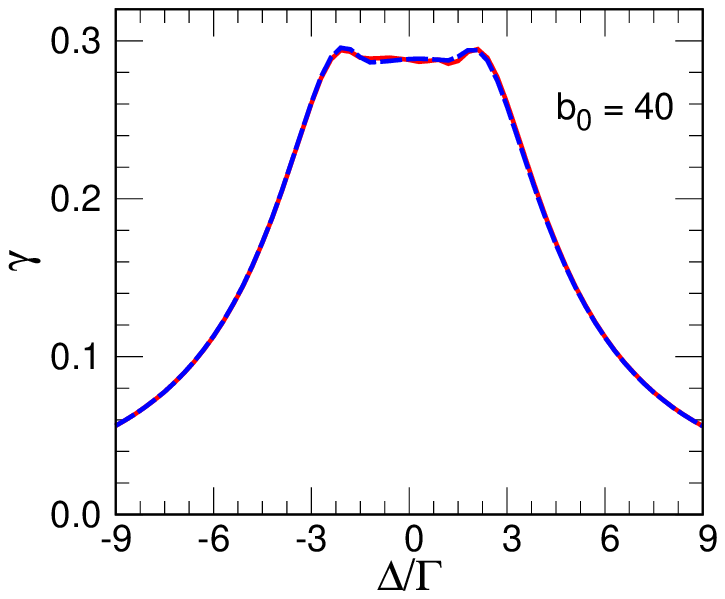}}
\caption{\label{FigFwdM}
The forward scattering rate (see text for description) versus 
detuning for different
cooperativity parameters, $b_0$. Each plot shows the results for
the full calculation with $N=2^{17}$ solid (red) line and
the model calculation long dash (blue) line. For all plots, the different
calculations are in such good agreement that the different lines
are indistinguishable. For the larger value of $b_{0}$ shown here, the scattered lineshape plateaus near-resonance, due to the fact that light does not penetrate the cloud in this regime.
}
\end{figure}

Figure~\ref{FigTotM} raises an interesting point.
References~\cite{scully2007,sutherland_2016_1,bromley_2016}
derived the broader line width Lorentzian (dashed orange line) by solving for the superradiant
time dependence uniformly excited across the Gaussian distribution of
atoms (timed-Dicke state). However, neither the paraxial nor eikonal approximation of a
continuum dielectric use this concept.
The width in Eq.~(\ref{EqMod}) is the single atom width $\Gamma$.
The larger width emerging from  Eq.~(\ref{EqPmod}) is solely due to
the interplay of the non-uniform light intensity across the atom
cloud as well as the phase change in
Eq.~(\ref{EqEf}). Figure~\ref{FigTotM} also shows that for larger values 
of $b_{0}$ \textemdash when the polarization ceases to significantly
penetrate the full cloud  \textemdash point dipole and continuous dielectric
models show a narrowing of the lineshape near resonance; the timed-Dicke
state models do not show this since the uniformly polarized state ansatz
becomes insufficient.

The calculations allow us to untangle the coherent scattering of photons
in the forward direction and the random scattering into large angles.
To obtain the forward scattering rate, we integrated Eq.~(\ref{EqSct})
over $\phi$ from 0 to $2\pi$ and integrated $\theta$ from 0 to
$\cos (\theta_{\rm max})=1 - 13.8/(k^2r_f^2)$. The $\theta_{\rm max}$
was chosen so that forward scattering has decreased by at least two
orders of magnitude from its maximum value. The result is shown in Fig.~\ref{FigFwdM}.
There is a plateau in the forward scattering rate
for a range of detuning around $\Delta =0$. The atomic calculations
of the forward scattering rate are, again, well
reproduced by the paraxial approximation of the continuum distribution
for all of the optical depths that were calculated. The eikonal
approximation also agreed well with the atom calculation for the
forward scattered photons except for at large $b_0$ and small $N$.

\begin{figure}
\resizebox{80mm}{!}{\includegraphics{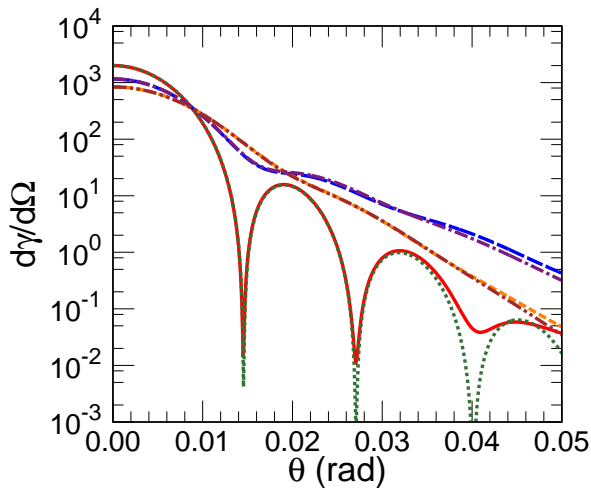}}
\caption{\label{FigFwdAngM}
The angular scattering rate per atom, Eq.~(\ref{EqSct}) for $b_0=40$
and $N=2^{17}$. There are calculations for 3 detunings: 0,  3/2, and $3\Gamma$.
The solid (red) and dotted (green) lines are the atom and paraxial
calculation for $\Delta =0$. The long dash (blue) and dash-dot (purple)
lines are the atom and paraxial calculation for $\Delta = 3\Gamma /2$.
The dashed (orange) and dash-dot-dot (brown) lines are the atom and
paraxial calculation for $\Delta = 3\Gamma$.
}
\end{figure}

Figure~\ref{FigFwdAngM} shows the full atom calculations as well as
the paraxial approximation for the angular scattering rate per atom,
Eq.~(\ref{EqSct}) for $b_0=40$ and $N=2^{17}$. This shows 
that coherently scattered light can be reproduced
by the continuum dielectric model. The figure shows results from three different detunings. The $\Delta =0$ case shows strong diffraction
minima due to the strong scattering of light in the center of the cloud;
these minima
are too deep for the continuum calculation because it does not include
the random scattering from pointlike atoms.
At large detuning, the absorption is less, so the scattered light 
more closely follows a Gaussian form.

\subsection{$\xi =2$, dilute gas limit}

\begin{figure}
\resizebox{80mm}{!}{\includegraphics{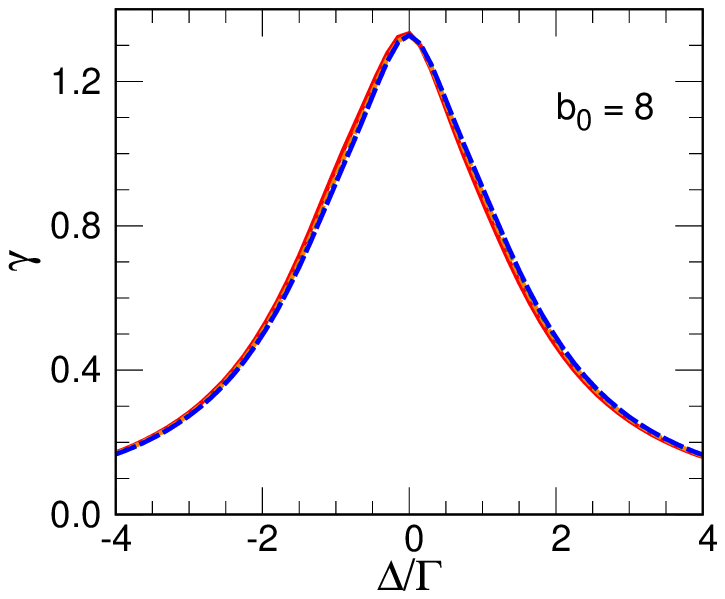}}
resizebox{80mm}{!}{\includegraphics{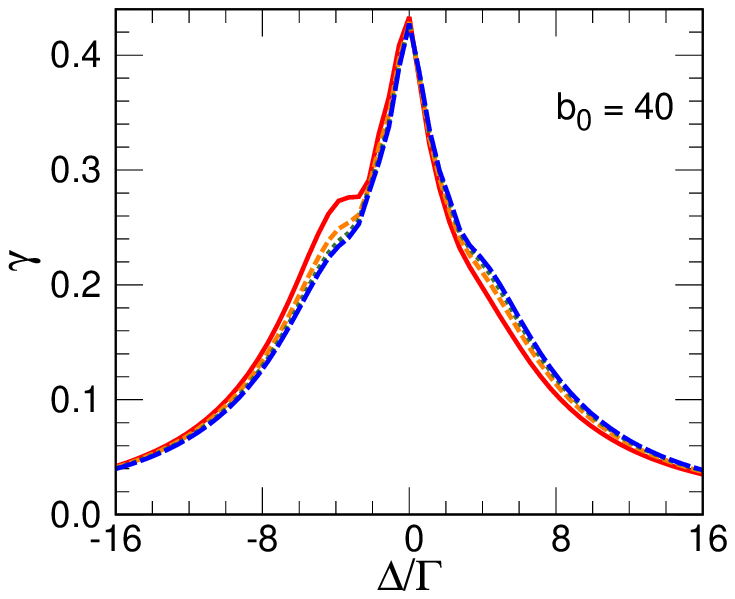}}
\caption{\label{FigTot2}
The total scattering rate per atom, Eq.~(\ref{EqSct}), versus
detuning; same as Fig.~\ref{FigTot} except for $\xi =2$.
Each plot shows the results for
four different numbers of atoms: solid (red) line $N=2^{11}$,
dashed (orange) line
is for $N=2^{13}$, the dotted (green) line is for $N=2^{15}$,
and long dashed (blue) line $N=2^{17}$.
}
\end{figure}

When the Gaussian cloud is elongated in the laser propagation direction, there
is more absorption and focusing of the laser beam.
For $\xi =2$, the total scattering versus detuning for $N=2^{11}$, $2^{13}$,
$2^{15}$, and  $2^{17}$ are shown
in Fig.~\ref{FigTot2}. Unlike Fig.~\ref{FigTot}, the calculations with
different numbers of atoms give different results for $b_0\geq 24$
indicating the breakdown of the eikonal approximation.
In all of the calculations, the scattering rates converge to a symmetric
form as $N\to\infty$, but, for smaller $N$, the scattering rate is larger
for $\Delta < 0$. In fact, there is a significant hump for $b_0=40$ at 
$\Delta <0$
for the $N=2^{11}$ calculation, solid (red) line.
These new parametrical dependencies arise due to focusing of the
light at $\Delta <0$, which correlates with a breakdown of the eikonal
approximation. The focusing bends the light skirting the edge of the cloud
so that it interacts more strongly with atoms to the back of the cloud
than would happen without focusing: since more light goes through
the cloud, there is more scattering.
The focusing is also the cause of the more extreme case of the effect
seen in Ref.~\cite{sutherland_2016_1} where atoms at the back of the cloud with
$\xi\gg 1$ were more strongly
excited than atoms at the front of the cloud. The effect
is larger at smaller $N$ because the cloud is smaller
and denser which leads to more focusing.
Even more than $\xi = 1$, the large $b_0$ scattering rate has a dependence
more similar to $|\Delta |$ than to $\Delta^2$.

\begin{figure}
\resizebox{80mm}{!}{\includegraphics{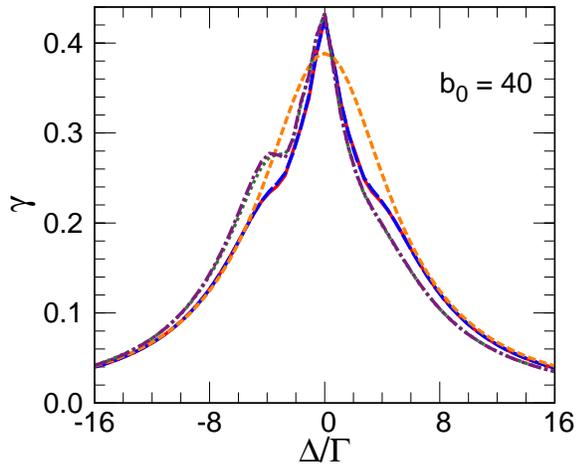}}
\caption{\label{FigTotM2}
The total scattering rate per atom, Eq.~(\ref{EqSct}), versus detuning;
same as Fig.~\ref{FigTotM} but for $\xi = 2$.
Each plot shows the results for
the full calculation, Eq.~(\ref{EqPart}), ($N=2^{17}$ solid (red) line
and $N=2^{11}$ dotted (green) line), for
the model calculation, Eq.~(\ref{EqMod}), ($N=2^{17}$ long dash (blue) 
line and $N=2^{11}$ dash-dot (purple) line), and
for a Lorentzian proportional to $1/(1+(2\Delta/\Gamma ')^2)$,
dashed (orange) line. Note there are 5
curves plotted.
}
\end{figure}

These distributions are well reproduced by the continuum dielectric
model that uses the paraxial approximation. Figure~\ref{FigTotM2} shows
the comparison between the atomic and the continuum model calculations  of
the total scattering rate for $N=2^{11}$ and $2^{17}$; the continuum
model calculations are nearly indistinguishable from the atomic
calculations. Note
that the hump at negative $\Delta$ for $N=2^{11}$ is well reproduced.

\begin{figure}
\resizebox{80mm}{!}{\includegraphics{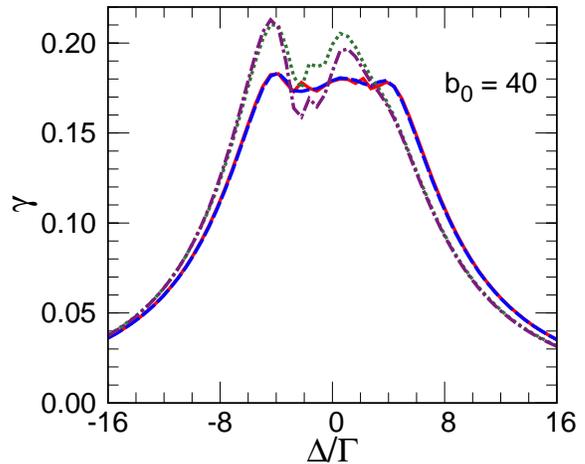}}
\caption{\label{FigFwdM2}
The forward scattering rate (see text for description) versus 
detuning; same as Fig.~\ref{FigFwdM} but for $\xi = 2$.
Each plot shows the results for
the full calculation ($N=2^{17}$ solid (red) line and
$N=2^{11}$ dotted (green) line) and
the continuum model calculation ($N=2^{17}$ long dash (blue) line and
$N=2^{11}$ dash-dot (purple) line).
The distributions
for other $b_0$ have better agreement.
}
\end{figure}

The forward scattering rate for $\xi =2$, $b_0=40$, and $N=2^{17}$
and $2^{11}$ are shown in Fig.~\ref{FigFwdM2}. This has a similar
form to the spherical cloud although the plateau starts at smaller $b_0$
(not shown). The results from the paraxial approximation to the continuum model
are also shown. The $N=2^{17}$ results are in excellent agreement but
there is a noticeable difference for $N=2^{11}$ and small $|\Delta|$.
The difference arises when the light diffracts back into the atom cloud.
We found that including the next order term did not improve the paraxial
approximation which suggests that this difference is due to a breakdown
of the continuum dielectric model for light propagation at higher density.
The possibility that this difference arises because we use the low
density form of the susceptibility, instead of Eq.~(\ref{EqChi}),
is addressed in the next section.

\subsection{Denser gases}\label{SecHD}

Reference~\cite{javanainen2014} showed that the Clausius-Mossotti 
(or Lorentz-Lorenz)
form of the susceptibility, Eq.~(\ref{EqChi}), does not describe the
model of light scattering from stationary atoms, Eq.~(\ref{EqPart}).
In all of the calculations above, the paraxial approximation of the
continuum model used the low density form for the susceptibility,
Eq.~(\ref{EqChild}). This did not make much difference in the
calculations because the maximum of $\chi_e^{(ld)}$ was not very
large. Nevertheless, we also found that the calculations using $\chi_e^{(ld)}$
were more accurate than using $\chi_e$ from Eq.~(\ref{EqChi})
for smaller $N$.

\begin{figure}
\resizebox{80mm}{!}{\includegraphics{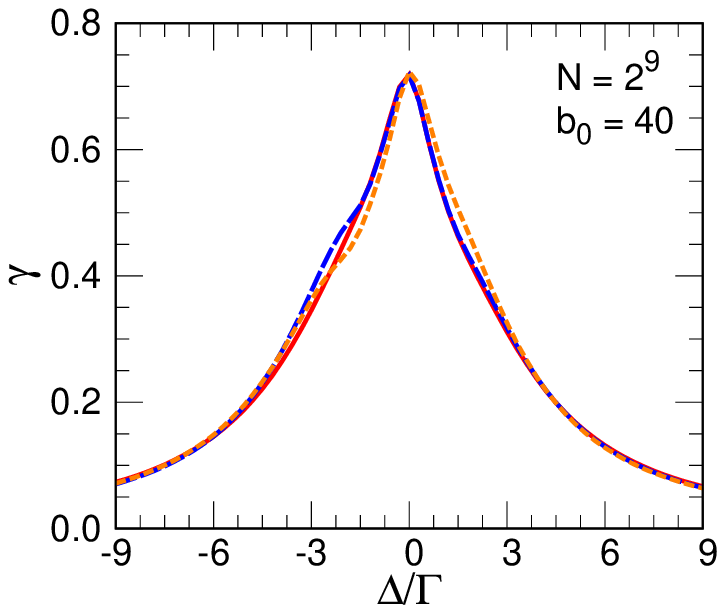}}
\resizebox{80mm}{!}{\includegraphics{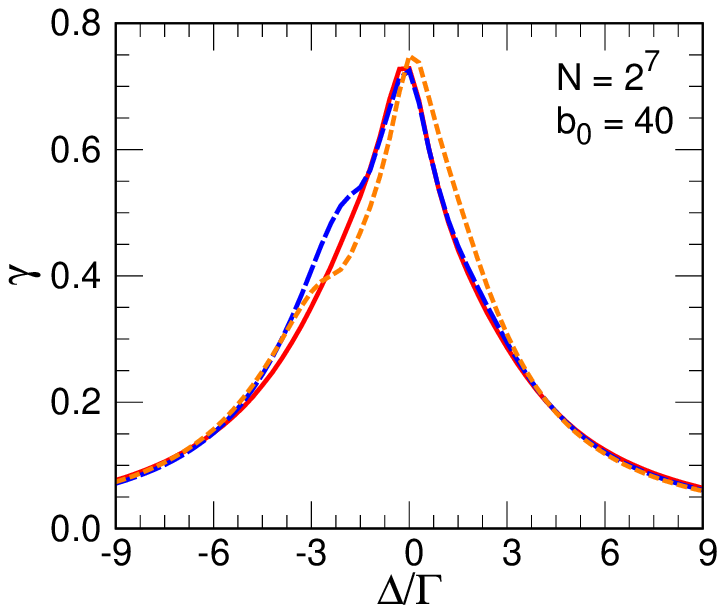}}
\caption{\label{FigTotHD}
The total scattering rate per atom, Eq.~(\ref{EqSct}), versus detuning
for $\xi = 1$; same as Fig.~\ref{FigTotM} but for $N=2^9$ and $2^7$.
Each plot shows the results for
the full calculation, Eq.~(\ref{EqPart}), solid (red) line,
the continuum model calculation, Eq.~(\ref{EqMod}), 
 using $\chi^{(ld)}_e$ long dash (blue) line, and
the continuum model using the Claussius-Mossotti
(or Lorentz-Lorenz) electric susceptibility, Eq.~(\ref{EqChi}),
dashed (orange) line.
}
\end{figure}

Figure~\ref{FigTotHD} shows a comparison between the atom calculation
(solid (red) line) and the paraxial equation using the low density electric
susceptibility (long dashed (blue) line), Eq.~(\ref{EqChild}), and the 
the Clausius-Mossotti susceptibility
(dashed (orange) line), Eq.~(\ref{EqChi}). The continuum calculation
using the low density electric susceptibility seems to overestimate
the effect from focusing for $\Delta\sim -3\Gamma$ whereas the
Clausius-Mossotti susceptibility is not accurate throughout the range
$|\Delta | < 3\Gamma$. We found
that the second order correction to the paraxial approximation did
not explain the difference for $N=2^9$; although the correction to the paraxial
approximation was not negligible for the $n=2^7$ calculation, it could
not explain the difference with the atom calculation.
Overall, the low density form of the susceptibility
gives a more accurate representation of the
total scattering versus detuning. The Clausius-Mossotti susceptibility gives a
blue shift to the line whereas the low density form gives a slight red
shift due to focusing for $\Delta <0$. The size of the effect is
smaller than Ref.~\cite{javanainen2014} because they used a much higher density
and they used a uniform density
whereas a Gaussian density was used in this paper.
The calculations in Ref.~\cite{javanainen2014} were for $\rho/k^3=2$ whereas
Fig.~\ref{FigTotHD} has $\rho_{max} /k^3 = 0.27$ for $N=2^7$ and
0.14 for $N=2^9$. These results show that the Clausius-Mossotti (or
Lorentz-Lorenz) electric susceptibility does not reproduce the
stationary atom calculation even in cases where
$\rho /k^3 < 1$.

\section{Conclusions}\label{SecCon}

We have performed calculations of light scattering from a weakly driven
Gaussian cloud of stationary atoms. We showed nontrivial effects on
the scattering when moving from small to large optical depths. We also demonstrated
a method that can solve for light scattering from many more atoms
than is typical in current calculations.
Thus, simulations can approach the number of atoms
in experiments; results for up to $N=2^{17}$ were presented.

We showed that the photon scattering rate
versus detuning is quite different from a Lorentzian at larger
optical depths. This is because when poralization begins to build in the front of the cloud, the on-resonant forward scattered light does not propagate through and plateaus. For larger numbers
of atoms, the total and forward scattering rates were quantitatively
reproduced by a continuum model that used the low density
expression for the electric susceptibility. Even though it only contains linear absorption, the continuum dielectric calculation gave better agreement with the full point dipole calculations than models that use the single photon superradiance framework. The full point dipole results for smaller atom number differ somewhat from the
continuum model. Interestingly, worse results were obtained
when using the Clausius-Mossotti (or Lorentz-Lorenz) form
for the electric susceptibility, in agreement with the
findings in Ref.~\cite{javanainen2014}.

We thank J. Ruostekoski for interesting discussions about this
concept and for suggesting the paraxial approximation as a method
for solving the continuum model.
FR was supported by the National Science Foundation under
Grant No.~1804026-PHY. Part of this work was performed under
the auspices of the U.S. Department of Energy by Lawrence
Livermore Laboratory under Contract DE-AC52-07NA27344. LLNL-JRNL-786838-DRAFT

\section{Appendix}

Below are more detailed description of numerical methods used in the
calculations above.

\subsection{Iterative method}\label{SecApp}

The steady state solution of Eq.~(\ref{EqPart}) involves the solution of
a linear equation. For most of the calculations in this paper, we restricted
the oscillators to only be in the $x$-direction which means only the terms
with $j=x$ are included. The discussion in this section will focus on this
case for simplicity but it should be clear how to generalize to include
all polarizations. For $N$ atoms, this leads to an $N\times N$
matrix equation of the form:
\begin{equation}\label{EqMat}
\sum_b A_{b'b}a_b = s_{b'}
\end{equation}
For small number of atoms (less than $\sim 1000$), we used Lapack
subroutines to directly solve for $a_b$. For larger number of atoms,
we used an iterative method based on successive over-relaxation.

The method proceeded in five stages. First, we ordered the atoms
in the direction of laser propagation; for the $\vec{k}=k\hat{e}_z$ used above, the
atoms are ordered from smallest $z$ to largest $z$; the $a_b$ are
updated in this order so that the atom with most negative $z$ is
updated first and the atom with most positive $z$ is updated last. Next, for
each atom $\beta$, we found the nearest $M-1$ atoms $b'$; these atoms will
have the largest $g$ (hence, the largest $A_{b'\beta }$). The third
step constructs an $M\times M$ linear problem using the $a_b$ from the
previous iteration. The smaller linear system is defined by
\begin{eqnarray}
\tilde{A}_{b'b}&=&A_{b'b}\; {\rm for}\quad b',b\in M\nonumber\\
 \tilde{s}_{b'} &=& s_{b'} - \sum_{b\notin M} A_{b'b}a_b\quad {\rm for}\; b'\in M
\end{eqnarray}
We next solve the much smaller linear equation
\begin{equation}\label{EqAtil}
\sum_b\tilde{A}_{b'b}\tilde{a}_b=\tilde{s}_{b'}
\end{equation}
using standard Lapack subroutines
and update only the atom $\beta$: $a_\beta = (a_\beta + \tilde{a}_\beta )/2$.

We order the atoms from small to large $z$ because atoms with smaller $z$
affect those at larger $z$ more strongly than vice versa. By taking
them in order, the convergence speed was improved. More importantly,
by directly solving Eq.~(\ref{EqAtil}), we are able to account for the large coupling between
close pairs (or triples or quadruples etc.) of atoms.

We were able to converge all of the calculations with more than
$2^{10}$ atoms using this method. Typically, we used 9 iterations
before convergence. Most of the calculations converged with $M-1=7$
closest atoms. The calculations with the largest $b_0$ and smallest
$N$ sometimes did not converge for $M=8$ but did converge for $M=16$.
The calculation speed improves with smaller $M$ so we first did all
calculations with $M=8$ and only repeated the failed ones with $M=16$.
The failed calculations were easy to determine because they had
discontinuous jumps in scattering rate versus detuning.

This algorithm was much faster than directly solving Eq.~(\ref{EqMat}).
This also solved the problem of memory ($A_{b'b}$ has
$N^2$ complex numbers); when $N$ was too large for the
memory of our computer, we could compute
the $A_{b'b}$ on the fly instead of storing them in an array.
Although not reported here, we did calculations with $N=2^{18}$
atoms and one test calculation with $N=2^{19}$. The calculation
with $2^{19}$ atoms would require an $A$ with $\sim 1/4$ trillion
elements (i.e. over a terabyte of RAM) if done by direct solution.
Such large $N$ 
can be reached because the density decreases with $N$ which allows
a smaller value of $M$ to be used. The algorithm
can be parallelized; most of the calculations were done on a 4 processor
workstation, but the largest calculations were done on a 20 processor
workstation.

\subsection{Paraxial approximation}\label{SecPar}

To derive the paraxial approximation as used in this paper, start from
the expression for the exact Maxwell equation, Eq.~(\ref{EqMax}), and
substitute the form
\begin{equation}
\vec{E}=E_0e^{ikz}\vec{\psi}\equiv E_0e^{ikz}(\vec{\psi}_T+\hat{z}\psi_z)
\end{equation}
where $\vec{a}_T\equiv a_x\hat{x}+a_y\hat{y}$ for any vector $\vec{a}_T$.
This gives
\begin{eqnarray}\label{EqPar1}
&i&\frac{\partial\vec{\psi}_T}{\partial z} + \frac{1}{2k}\nabla_T^2\vec{\psi}_T
+\frac{k}{2}\underline{\chi}_e\vec{\psi}_T\nonumber \\
&=&  
-\frac{1}{2k}\frac{\partial^{2}\vec{\psi}_T}{\partial z^2} +\frac{1}{2k}
\vec{\nabla}_T 
\left( \vec{\nabla }_T\cdot\vec{\psi}_T+ik\psi_z+
\frac{\partial \psi_z}{\partial z}\right)
\end{eqnarray}
and
\begin{equation}\label{EqPar2}
\psi_z = \frac{i}{k}\vec{\nabla}_T\cdot\vec{\psi}_T-\frac{1}{k^2}
\nabla_T^2\psi_z+\frac{1}{k^2}\frac{\partial}{\partial z}
\vec{\nabla}_T\cdot\vec{\psi}_T
\end{equation}
where $\underline{\chi}_e\vec{\psi}_T\equiv\chi_e\psi_x \hat{x}$.
These equations are exact but are still difficult to solve. To simplify
the equations below we define the operator ${\cal B}$
as
\begin{equation}
    {\cal B}\vec{\psi}_T \equiv
    i\frac{\partial\vec{\psi}_T}{\partial z} + \frac{1}{2k}\nabla_T^2\vec{\psi}_T
+\frac{k}{2}\underline{\chi}_e\vec{\psi}_T
\end{equation}
We modify
Eq.~(\ref{EqPar1}) by substituting Eq.~(\ref{EqPar2}) for $\psi_z$ to
give
\begin{eqnarray}\label{EqPar3}
\frac{1}{k}{\cal B}\vec{\psi}_T
&=&
-\frac{1}{2k^2}\frac{\partial^{2}\vec{\psi}_T}{\partial z^2}\nonumber \\
&+&\frac{1}{2k^2}
\vec{\nabla}_T 
\left( -\frac{i}{k}\nabla_T^2\psi_z
+\frac{i}{k}\frac{\partial}{\partial z}
\vec{\nabla}_T\cdot\vec{\psi}_T+
\frac{\partial \psi_z}{\partial z}\right) \nonumber \\
\end{eqnarray}

To obtain the paraxial approximation, one scales $x,y$ by a width
${\rm w}$ and $z$ by a length $L\equiv {\rm w}/f$ 
(i.e. $x={\rm w}\bar{x}$, $y={\rm w}\bar{y}$,
and $z=L\bar{z}$ with the barred coordinates being dimensionless). The ratio
${\rm w}/L = f$ is set equal to $f\equiv 1/(k{\rm w})$. For the paraxial
approximation, $f$ should be small which means the distance scale of
variations in $x,y$ should be large compared to $1/k$ and the distance
scale of variations in $z$ should be large compared to that in $x,y$.
Substituting this scaling into the differential equations suggests
that the three terms on the right hand side of Eq.~(\ref{EqPar2}) are of
order $f^1$, $f^2$, and $f^3$ respectively. The terms on the
left hand side of Eq.~(\ref{EqPar3}) are of order $f^0$ or $f^2$
and on the right hand side are of order $f^3$ if they involve $\psi_z$
and $f^4$ if they involve $\vec{\psi}_T$.

The functions are written as a series
\begin{eqnarray}
\vec{\psi}_T&=&\vec{\psi}_T^{(0)}+\vec{\psi}_T^{(2)}+\vec{\psi}_T^{(4)}+...
\nonumber \\
\psi_z&=&\psi_z^{(1)} +\psi_z^{(3)} +\psi_z^{(5)} +...
\end{eqnarray}
To obtain the equations for the different terms, one groups the
same orders together. For example, since the first term on the right
hand side of Eq.~(\ref{EqPar2}) is of
order $f^1$, a term like $(i/k)\vec{\nabla}_T\cdot\vec{\psi}_T^{(4)}$
is of order 5 since the $\vec{\psi}_T$ is order 4 and the operation
is of order 1. Equation~(\ref{EqPar2}) is transformed to
\begin{equation}\label{EqPar6}
\psi_z^{(n)} = \frac{i}{k}\vec{\nabla}_T\cdot\vec{\psi}_T^{(n-1)}-\frac{1}{k^2}
\nabla_T^2\psi_z^{(n-2)}+\frac{1}{k^2}\frac{\partial}{\partial z}
\vec{\nabla}_T\cdot\vec{\psi}_T^{(n-3)}
\end{equation}
when grouping terms of order $n$. Defining the order of the
$(1/k){\cal B}$ operator is somewhat problematic due to the
$\chi_e$ term. We take it to be an order 2 operator, consistent
with the two differential terms. Equation~(\ref{EqPar3}) is
transformed to
\begin{eqnarray}\label{EqPar4}
\frac{1}{k}{\cal B}\vec{\psi}_T^{(n)}
=
&-&\frac{1}{2k^2}\frac{\partial^{2}\vec{\psi}_T^{(n-2)}}{\partial z^2}
+\frac{1}{2k^2}
\vec{\nabla}_T 
( -\frac{i}{k}\nabla_T^2\psi_z^{(n-1)}\nonumber \\
&+&
\frac{i}{k}\frac{\partial}{\partial z}
\vec{\nabla}_T\cdot\vec{\psi}_T^{(n-2)}+
\frac{\partial \psi_z^{(n-1)}}{\partial z})
\end{eqnarray}
when grouping terms of order $n+2$.
If $n=0$, 1, or 2, the functions on the right hand side can have
negative superscript. The rule for evaluating these are:
any function with a negative superscript is zero everywhere.

The case discussed in the paper has the atoms only being polarizable in
the $x$-direction. This means all of the quantities of interest can
be calculated from the $\psi_x$. Thus, through order 2, the equations
to be solved are
\begin{eqnarray}
{\cal B}\hat{x}\psi_x^{(0)}
&=& 0\qquad \psi_y^{(0)}=0\nonumber \\
\psi_z^{(1)} &=& \frac{i}{k}\frac{\partial \psi_x^{(0)}}{\partial x}\nonumber \\
{\cal B}\hat{x}\psi_x^{(2)}
&=& -\frac{1}{2k}\left(\frac{\partial^2\psi_x^{(0)}}{\partial z^2} 
+\frac{\partial^2[\chi_e\psi_x^{(0)}]}{\partial x^2}\right)\hat{x}
\end{eqnarray}
with the $\psi_x^{(2)}$ set to 0 as $z\to -\infty$. Note the
$\psi_y^{(2)}$ is nonzero, but it is not used in our calculations.

\bibliography{gausscloud}

\end{document}